\documentclass[namedreferences]{solarphysics}
\usepackage[hyperref,optionalrh,solaromanenum]{spr-sola-addons} % For Solar Physics 
\usepackage{graphicx}                    % For eps figures, newer & more powerfull
\usepackage{amssymb}                    % useful mathematical symbols
\usepackage{color}                       % For color text: \color command
\usepackage{breakurl}                         % For breaking URLs easily trough lines
\usepackage{amsbsy}

\chardef\us=`\_

\newcommand{\vect}[1]{\boldsymbol{#1}}

\begin{document}

\begin{article}

\begin{opening}

\title{Probing the Sunspot Atmosphere with Three-Minute Oscillations}

%%%%%%%%%%%%%%%%%%%%%%%%%%%%%%%%%%%%%%%%%%%%%%%%%%%
%% Authors Names
%

\author[addressref={istp},corref,email={deres@iszf.irk.ru}]{\inits{A.S.}\fnm{Anastasiia}\lnm{~Deres}\orcid{0000-0003-0186-062X}}
\author[addressref={istp},corref]{\inits{S.A.}\fnm{Sergey}\lnm{~Anfinogentov}\orcid{0000-0002-1107-7420}}

%%%%%%%%%%%%%%%%%%%%%%%%%%%%%%%%%%%%%%%%%%%%%%%%%%%
%% Runningheads
%
\runningauthor{A.S. Deres, S.A. Anfinogentov} 
\runningtitle{Probing the Sunspot Atmosphere with Three-Minute Oscillations}

%%%%%%%%%%%%%%%%%%%%%%%%%%%%%%%%%%%%%%%%%%%%%%%%%%%
%% Affilations 
%% id shold be the same with \author addressref value.
%\address[id={}]{}
\address[id={istp}]{Institute of Solar Terrestrial Physics of the Russian Academy of Sciences,  Lermontov st. 126a, Irkutsk 664033, Russia}
%%%%%%%%%%%%%%%%%%%%%%%%%%%%%%%%%%%%%%%%%%%%%%%%%%%
%%% Abstract 

\begin{abstract}
	We present a seismological method to probe the solar atmosphere above sunspot umbrae with three-minute oscillations.
	Our technique allows us to estimate both the vertical distance between atmospheric layers and the wave-propagation speed without specifying any extra parameters, in particular, the phase speed of the wave or the emission formation heights.
	Our method uses the projected wave paths of slow MHD waves  propagating through the atmospheric layers of different heights and guided by the magnetic field. The length of the projected wave path depends upon the distance between the layers and the  inclination angle of the magnetic field with respect to the line of sight, allowing us to estimate the distance between the layers from measured  projected wave paths and the local magnetic-field vector.  In turn, the wave-propagation delay registered at different heights allows for the calculation of the phase speed.
	We estimated the vertical distance between the emission layers at the temperature minimum (1600\,\AA) and transition region (304\,\AA), as well as the average phase speed above the sunspot umbrae, for three active regions.
	We found that the distance between the 1600\,\AA\ emission layer and the transition region above the sunspot umbrae lies in the range of 500 -- 800\,km. The average phase speed between these layers was found to be about 30\,km\,s$^{-1}$ giving the sound speed of 6\,km\,s$^{-1}$. The temperature between the layers has been roughly estimated as 3000\,K and corresponds to the region of the temperature minimum.
	The results obtained are consistent with the semiempirical model of the sunspot umbrae atmosphere by  \cite{fontenla_semiempirical_2009}.
\end{abstract}

%%%%%%%%%%%%%%%%%%%%%%%%%%%%%%%%%%%%%%%%%%%%%%%%%%%
%% Keywords
%

\end{opening}
%-------------------------------------------------

%%%%%%%%%%%%%%%%%%%%%%%%%%%%%%%%%%%%%%%%%%%%%%%%%%%
%% Sections
%
\section{Introduction}
  \label{s:intro} 
  
 Investigation of waves and oscillations in the solar atmosphere is a very important part of  solar physics. Oscillations are natural probes carrying information about  the medium where they propagate. Therefore oscillations and waves can be used to measure parameters of the plasma in the solar atmosphere \citep{1983SoPh...82..369Z, 2012RSPTA.370.3193D}. This is the subject of coronal seismology, which is a new branch of solar physics. 
 
 Oscillations in the sunspot atmosphere were  discovered in  optical observations more than 40 years ago \citep{1969SoPh....7..351B,1972SoPh...27...71G}. Since that time, they have been intensively investigated by many researchers. Detailed information on this topic can be found in numerous reviews \citep[\textit{e.g.}][]{1992ASIC..375..261L,2000SoPh..192..373B, 2006RSPTA.364..313B, 2008sust.book.....T,2015STP.....1b...3S,2016GMS...216..467S}.
 
 The great  interest in sunspot oscillations is connected  with the opportunity of applying them as  seismological diagnostics of the  solar atmosphere above sunspots.  In particular, observations of the acoustic cut-off frequency allow one  to estimate the inclination of the magnetic-field. \cite{2014A&A...561A..19Y}  used observations of the acoustic cut-off frequency at several levels of  the sunspot atmosphere to estimate  magnetic field inclination angles.  
 \citet{2017ApJ...837L..11C} used the three-minute oscillations for the estimation of the Alfv\'en speed, parameter $\beta$, and mass density within the umbra.

 Additional information can be obtained from observations of sunspot oscillations simultaneously at several heights in the solar atmosphere. Multilevel observations aimed at measuring phase delays started in the 1970 \citep{1978SoPh...58..347G, 1983A&A...123..263U, 1985ApJ...294..682L}. It was found that the three-minute oscillations are upward-propagating slow magneto-acoustic waves. This finding is consistent with theoretical modelling of this phenomenon 
\citep[\textit{e.g.}][]{2011ApJ...728...84B, 2012A&A...539A..23S, 2015ApJ...808..118C,  2015A&A...580A.107S}.
    
 New observational capabilities reinforced the interest in the three-minute oscillations, leading to a number of studies dedicated to measuring and interpreting delays between oscillations observed at the different levels of the sunspot atmosphere. Measurement of the delays together with the information about the emission formation heights  enables the estimation of the wave-propagation phase speed. 
    
 \cite{2013SoPh..284..379K} measured the phase delays of the three-minure oscillations observed  in different spectral lines. They found that the observed delays correspond to a propagation speed that is significantly  higher than the expected sound speed in the corresponding atmospheric layers. 
 \cite{reznikova_three-minute_2012} measured the phase delays between the layers visible in different extreme ultraviolet (EUV) channels of the \textit{Atmospheric Imaging Assembly} (AIA) onboard the \textit{Solar Dynamics Observatory} (SDO).
 The measured delay between the signals in the 1600\,\AA\ and 304\,\AA\  channels gave a phase speed speed about 70\,km\,s$^{-1}$,  which is higher than the expected sound speed in the chromosphere (10 -- 30 \,km\,s$^{-1}$).
 The delay between the 304\,\AA\ and 171\,\AA\ signals corresponds to a  speed of 83\,km\,s$^{-1}$.
 \cite{reznikova_three-minute_2012} argue  that this value is realistic for the sound speed, because it corresponds to a temperature of $2\times 10^5$ K, which is very close to the expected temperature of the oscillating  layers. 
    
 The partial inconsistency of the measured speed of the slow  magnetoacoustic waves and the expected sound speed in the sunspot atmosphere may be due to the difference between the real emission formation heights and the heights given by the model of the atmosphere.
 For example, \cite{2013SoPh..284..379K}  pointed out that the interpretation of the measured delays swap on the spectral-lines' formation heights given by the models of the solar atmosphere.
 It is important  that these heights are different in different models.
 We should also note that at the level of the temperature minimum and lower chromosphere, the main oscillation period (three minutes) is of the same order of magnitude as the acoustic cut-off frequency. Hence, the phase speed of the slow waves in those layers can be significantly higher than the sound speed, due to the dispersion caused by the density stratification.

 \cite{2015ARep...59..959D}  assumed that the wave-propagation speed is known and is equal to the local sound speed.
 This assumption is applicable to an acoustic wave in the upper chromosphere and the corona, where the  acoustic cut-off frequency is lower than the oscillation frequency.
 Therefore, relative heights of the emission layers can be calculated from the measured delays. For the  active regions  investigated, the  distance between the temperature minimum (1600\,\AA) and the transition region (304\,\AA)  was found to be less than 1 Mm. We compared the values obtained with two different empirical models of the sunspot umbra atmosphere. The first one is the model of \cite{maltby_new_1986} that is widely used.  The second one is a recent model developed by \cite{fontenla_semiempirical_2009}.
 
 The above models give very different  temperature dependences upon the height. 
 The modern model has a sharp temperature increase at a height about 1000 km above  the photosphere, where the temperature rises from about 4000~K up to the coronal values. In fact, this model does not have an extended chromosphere at all. On the other hand, the model of \cite{maltby_new_1986} has a pronounced 1000\,km wide chromospheric plateau, where the temperature changes relatively slowly. The wave-propagation delays measured by \cite{2015ARep...59..959D} support  the modern  model of \cite{fontenla_semiempirical_2009}.
 
 The aim of this work is to measure  both the distance and  average phase speed between different atmospheric layers from the observational data only, without making any assumptions about the phase speed and spectral-line formation heights. Consideration of the 3D geometry of the propagating wave is needed for this purpose. This knowledge allows us to  measure the distance between the locations where the wave passes through different atmospheric layers. These measurements are combined with  the magnetic-field vector extrapolated from the photospheric observations, giving us the estimate of the distance between the layers. The phase speed is then calculated from the estimated distance and measured phase delay.
  
 In our analysis, we assume that the three-minute oscillations propagate along the magnetic-field lines, because they are known to be slow magnetoacoustic waves, and the magnetic effects \citep{2015A&A...582A..57A}  are neglected, as the plasma-$\beta$ in sunspots is low. Another assumption is  that the layers where the 304\,\AA\ and 1600\,\AA\ emission is formed  are thin in comparison with the wave length of the propagating wave.    
    
 The seismological inversion method allowing for estimation of  the average phase speed and distance between two atmospheric layers is described in Section~\ref{sec: method}. In Section~\ref{sec: application}, the method is illustrated with the analysis of the three-minute oscillations observed with SDO/AIA at 1600\,\AA\ and 304\,\AA\  wavelengths in three active regions. The conclusion is given in Section~\ref{sec:conclusion}.

 \section{Description of the Method}
 
 \label{sec: method}
 
The proposed method uses imaging observations of an upward-travelling slow magnetoacoustic wave at two different levels, which allows us to estimate the average wave-propagation phase speed and distance between these levels. 
As an input, our algorithm needs two image sequences of the sunspot taken at different heights and the map of the magnetic-field vector at the height roughly corresponding to the lowest height. The latter can be either a photospheric vector  magnetogram, or results of the field extrapolation from a photospheric line of sight (LOS) magnetogram.

In a general case, a wave propagates along the magnetic field obliquely to the LOS (\textit{e.g.} \cite{2015ApJ...802...45C}).
Therefore, it appears at different wavelengths in different positions in the plane of the sky.
The distance  between these positions is the projection of the wave path  onto the plane of sky   and  can be measured together with the propagation time.
In combination with the knowledge of the wave-propagation direction, the phase delay and the wave path projection allow us to calculate the distance between the line formation layers and the average phase speed as well.
These quantities  can be measured for all pixels for further estimation of average values and uncertainties.

    \subsection{Measuring Delays and Wave Path Projections}
     
\begin{figure} 
	\centerline{\includegraphics[width=0.5\linewidth]{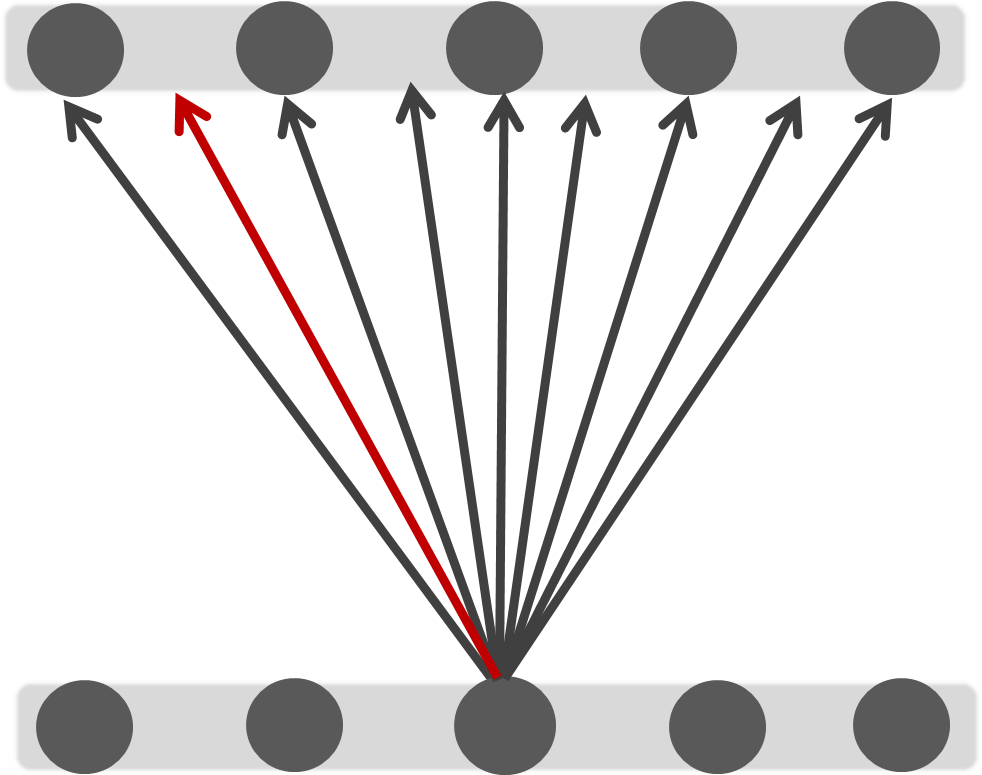}}
	\caption{Possible paths of the wave propagation (arrows) from the lower emission layer to the upper one. The distinct image pixels are indicated by circles. We assume that the wave  propagates along the path where the correlation coefficient between the signals at the lower and upper layers reaches its maximum (red arrow)}
	\label{fig:fig3}
\end{figure}

    Wave phase delays and wave path projections are measured independently for every pixel at the lower emission level.
    First, we fix one starting pixel at the lower  layer (see Figure \ref{fig:fig3}) and compute the cross-correlation function  of the signals at the selected pixel at the lower level and the neighbouring pixels at the upper level.
    Thus, we get a phase delay and a correlation coefficient for every pair of pixels.
    Then, we find the position at the upper layer where the cross-correlation coefficient reaches its maximum.  To estimate the maximum position with the subpixel accuracy, cubic interpolation is used.
    The straight line connecting the selected pixel centre at the lower layer and  the position of the correlation coefficient maximum at the upper layer gives us the wave path in the 3D space.  Since the distance between the layers is unknown at this stage, we can measure only the projection of the wave path onto the plane of the sky.
    
    Repeating this procedure for all pixels at the lower layer, we obtain maps of the following parameters: components of the wave path projected onto the picture plane $\Delta_x$ and $\Delta_y$, phase delay [$\tau$], and the cross-correlation coefficient between the signals at the lower and
    upper layers.  
    For the subsequent analysis,  we select only the pixels where the cross-correlation coefficient is greater than~0.6.
     
\begin{figure}
        	\centerline{\includegraphics[width=1\textwidth,clip=]{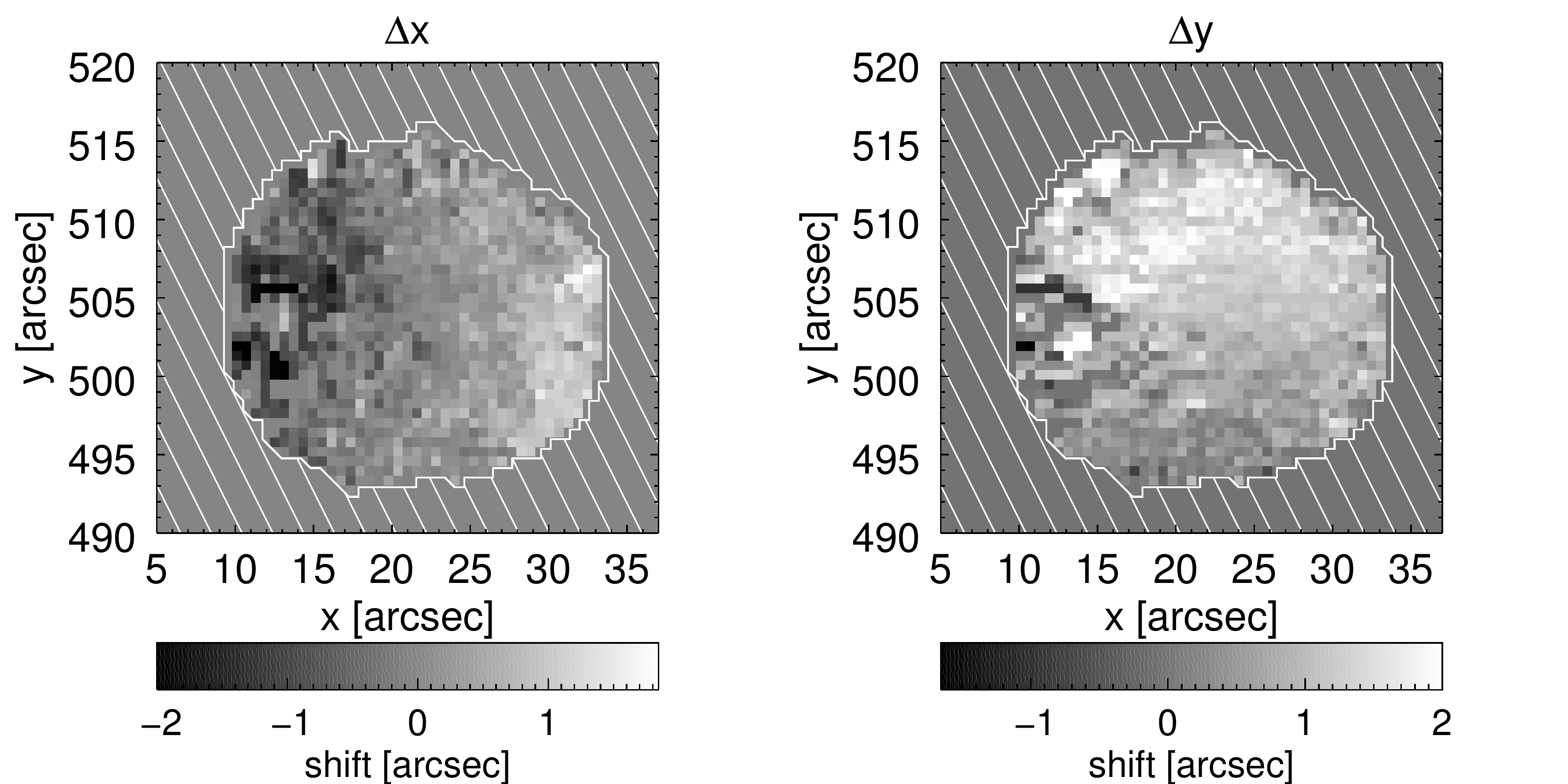}}
        	\caption{The maps of the $\Delta_x$ (left panel) and $\Delta_y$ (right panel) components of the wave paths projected onto the plane of the sky for the wave propagation between the levels observed in the channels 1600\,\AA\ and 304\,\AA}
        	\label{fig:fig4}
        \end{figure}

    Figure \ref{fig:fig4} provides an example of  wave-path projection maps [$\Delta_x$ and $\Delta_y$]  calculated from the observations of the three-minute oscillation at 1600\,\AA\ and 304\,\AA\ wavelengths (see Section \ref{sec: application}).
    The dark-bright gradient that is well seen in both of the panels in Figure \ref{fig:fig4}  indicates that the wave is expanding with height.

\subsection{Estimating the Distance between Emission Layers and the Phase Speed}

	The projected wave paths and the phase delays depend on the magnetic-field geometry and the distance between the layers, visible at different wavelengths. In the sunspot atmosphere the magnetic field is expanding with height. Thus, the upward-propagating wave front also should expand, following the hosting magnetic flux tube.	
  
   	 The projection of the wave path onto the picture plane  $\Delta = \sqrt{\Delta_x^2 + \Delta_y^2}$    depends on the angle $\phi_z$ between the wave-propagation direction and the LOS ($\vect{z}$ axis):
   	 
   	\begin{equation}
   	  \Delta = l\sin{\phi_z},
   	\end{equation} where $l$ is the length of the actual wave path between these two layers.
   	
   	Since the three-minute oscillations are slow magnetoacoustic waves, they are guided by the magnetic field.   	
   	Therefore, the angle $\alpha$ could be derived from the  magnetic-field [$\vect{B}$] extrapolated from the photospheric measurements to the height of the lower layer ($h_1$ = 500 km was used):

  \begin{eqnarray}
	  \cos\phi_z &=& \frac{B_{z}}{B},\nonumber\\
	  \phi_z &=&  \arccos{\frac{B_{z}}{B},}
  \end{eqnarray}
 where $B_{z}$ is the LOS component of the magnetic-field [$\vect{B}$], and $B$ is its absolute value.
       	
\begin{figure} 
     	\centerline{\includegraphics[width=1.05\textwidth,clip=]{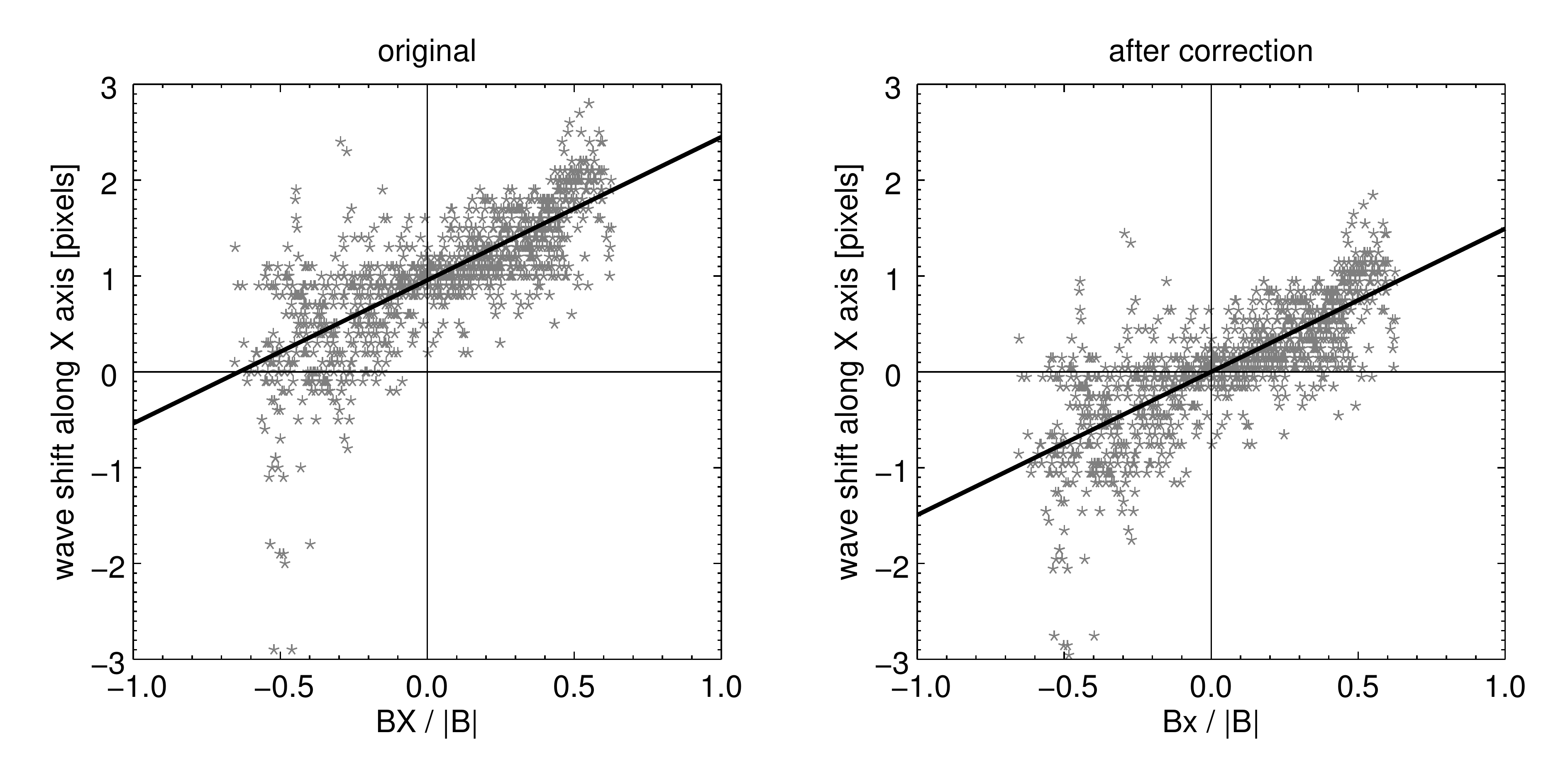}}
     	\caption{Dependence of the projected wave path upon the magnetic-field inclination in the active region NOAA 11131, wavelength 1600 and 304\,\AA. The original dependence is presented on the left panel while the right panel shows the result of the correction.}
     	\label{fig:fig5}
     \end{figure}

 In the areas where the  magnetic-field vector is parallel to the LOS ($\alpha = 0$), the wave path projection onto the plane of the sky $\Delta$ also must be equal to $0$. The left panel of Figure \ref{fig:fig5} shows an example of the dependence of the $\Delta_x$ wave path component on $B_x/B$ measured  from the observations of the three-minute oscillation at the temperature minimum and transition region (see Section \ref{sec: application}).
$B_x/B$ is the cosine of the angle $\phi_x$ between the magnetic-field vector [$\vect{B}$]  and \textbf{$\vect{x}$}  axis.
One can see that the wave path component $\Delta_x$ is not equal to 0 when cos$(\phi_x) = 0$.
This is an observational bias caused by non-ideal co-alignment of the images taken in different SDO/AIA channels.
This observational bias should be subtracted from the measured wave path components before further calculations.
The value of the bias is derived using a linear  least absolute deviation fit (\textsf{LADFIT} routine from the standard IDL library).
The fitting result is indicated by the straight blue line in Fig. \ref{fig:fig5}.
The right panel of Figure \ref{fig:fig3} shows the corrected data with the subtracted bias.
     
   The measured displacements [$\Delta_x$, $\Delta_y$] and wave-propagation time [$\tau$] allow us to calculate the wave-propagation distance 
 
\begin{equation}
    l=\frac{ \sqrt{\Delta_x^2 + \Delta_y^2}}{\sin\phi_z}
\end{equation} and the phase speed

\begin{equation} \label{eq:phase_speed}
 V=\frac{l}{\tau}.
\end{equation}
   
   Since the wave-propagation direction is not always perpendicular to the solar surface, the wave-propagation distance [$l$] and the vertical distance [$h$] between the emission layers are not of the same value.
   To find [$h$], we multiply [$l$] by the cosine of the angle $\psi$ between the magnetic field vector [$\vect{B}$] and the normal [$\vect{\hat{n}}$] to the solar surface.
   
     \begin{eqnarray}
   	  h&=&l\cos\psi,\\
   	  \cos\psi&=&\frac{n_{x}B_{x}+n_{y}B_{y}+n_{z}B_{z}}{B}.
     \end{eqnarray}
 In this way, the phase speed and distance between the emission layers are calculated for each pixel.

\begin{figure} 
	\centerline{\includegraphics[width=1\textwidth,clip=]{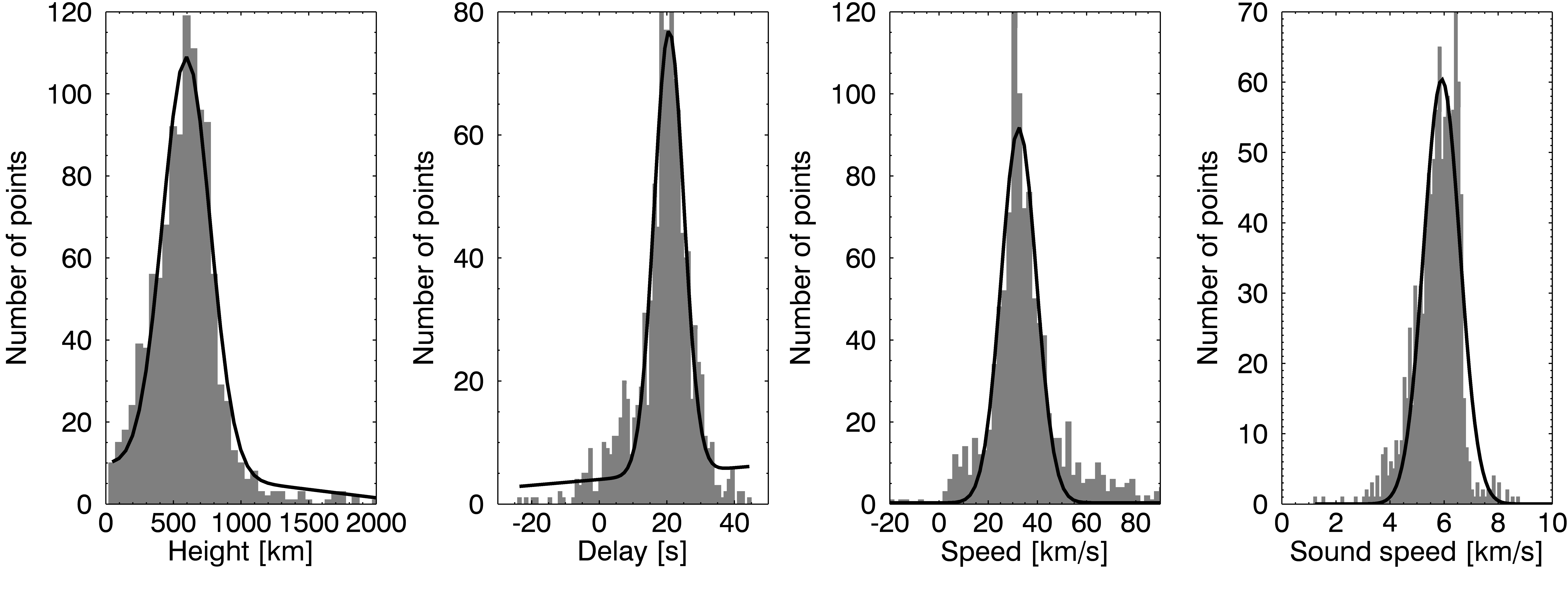}}
	\caption {Histograms of the vertical distance (left to right) between the emission layers, phase delay, phase speed,  phase delay, and sound speed. The Gaussian functions fitted to the histograms are shown with the black line.}  
	\label{fig:fig6}
    \end{figure}

The average value of these quantities and the confidence intervals are than calculated by fitting a Gaussian function to the histograms of the measured parameters (e.g. Figure \ref{fig:fig6}). 
The average values  are estimated  as maxima of the Gaussian functions fitted to the  histograms, while the width of the Gaussian gives us the confidence intervals. 

\subsection{Estimation of the Sound Speed}

Let us use the dispersion equation for magneto-acoustic-gravity waves in the low $\beta$ assumption to estimate the sound speed from the measured parameters:
 \begin{equation} \label{eq:disp_equation}
 \omega^2 = k^2c_{\rm s}^2 + \omega_0^2,
\end{equation}
where $\omega_0$ is the acoustic cut-off frequency
\begin{equation} \label{eq:acoust_cut-off}
 \omega_0 = \frac{g_{0} \gamma}{2 c_{\rm s}} \cos{\left (\alpha \right )}
\end{equation}
with the surface gravity of the Sun $g_0 = 0.274\ \mathrm{km\ s^{-2}}$,  inclination [$\alpha$] of the wave guide (i.e. magnetic field) with respect to the vertical direction, and adiabatic index $\gamma = \frac{5}{3}$.
The phase speed can be derived from Equation \ref{eq:disp_equation} as follows

\begin{equation} \label{eq:disp_phase_speed}
 v_{\rm p} \equiv \frac{\omega}{k} = \frac{c_{\rm s}}{\sqrt{1 -(\omega_0/\omega)^2}}.
\end{equation}

Substituting Equation \ref{eq:acoust_cut-off} in Equation \ref{eq:disp_phase_speed}, we get an equation for the sound speed [$c_{\rm s}$] with two positive solutions:
\begin{equation} \label{eq:sound_speed_solution}
c_{\rm s} = \frac{\sqrt{2}}{2} \sqrt{v_{\rm p}^{2} \pm \frac{v_{\rm p}}{\omega} \sqrt{- g_{0}^{2} \gamma^{2} \cos^{2}{\left (\alpha \right )} + \omega^{2} v_{\rm p}^{2}}}.
\end{equation}

The method developed allows us to measure only the average phase speed.
Therefore the sound speed estimated by Equation \ref{eq:sound_speed_solution} is an average measure and should be interpreted with caution.
Moreover, it is an estimation of the sound speed in the region with the \textit{lowest phase speed}.
According to Equation \ref{eq:disp_phase_speed}, we expect this condition to be satisfied in the lower chromosphere where the acoustic cut-off frequency [$\omega_0$] became significantly different from the frequency of the three-minute oscillations $\omega \gg \omega_0$, while the sound speed is still low.

Using $\alpha=0$ (vertical propagation), the typical oscillation frequency $\omega = \frac{2\pi}{180 {\rm s}}$, and the average value of the measured phase speed $v_{\rm p} = 30$\,km\,s$^{-1}$ (see Table \ref{tbl:1}), we obtain two possible estimations for the sound speed: 29.2\,km\,s$^{-1}$ (\lq\lq$+$\rq\rq\ in Equation \ref{eq:sound_speed_solution}) and 6.7\,km\,s$^{-1}$ (\lq\lq$-$\rq\rq\  in Equation \ref{eq:sound_speed_solution}).

The  \lq\lq$-$\rq\rq\ solution ($c_{\rm s} = 6.7$\,km\,s$^{-1}$) corresponds to the case of $\omega \gtrsim \omega_0$ and gives us a temperature about 4000\,K (assuming  weakly ionised plasma with the average particle weight of $\mu =1.27m_{\rm H}$ and $\gamma = 5/3$).
This temperature corresponds to the temperature minimum, which is a rather broad layer in the sunspot umbrae atmosphere (see Figure \ref{fig:models}) and can be seen in the 1600\,\AA\  SDO/AIA channel.
In the temperature minimum, the three-minutes oscillations are highly dispersive and have a phase speed essentially higher (up to infinity) than the local sound speed.

The  \lq\lq$+$\rq\rq\  solution for the sound speed in Equation \ref{eq:sound_speed_solution} gives us a value of $c_s \approx 30$\,km\,s$^{-1}$ that corresponds to the temperature of about 40,000\,K (assuming highly ionised plasma with particle weight of $\mu =0.62\,m_{\rm H}$ and $\gamma = 5/3$). This temperature can be associated with the transition region between the chromosphere and the corona.
Due to the high temperature and narrowness of this layer, it cannot make a significant contribution to the measured phase shift of the three-minute oscillations.
Therefore, this solution is not realistic and, in the subsequent analysis, we calculate only the \lq\lq$-$\rq\rq \ value given by Equation \ref{eq:sound_speed_solution}.

\section{Application to SDO/AIA Observations}

\label{sec: application}

\begin{figure} 
  	\centerline{\includegraphics[width=0.75\textwidth,clip=]{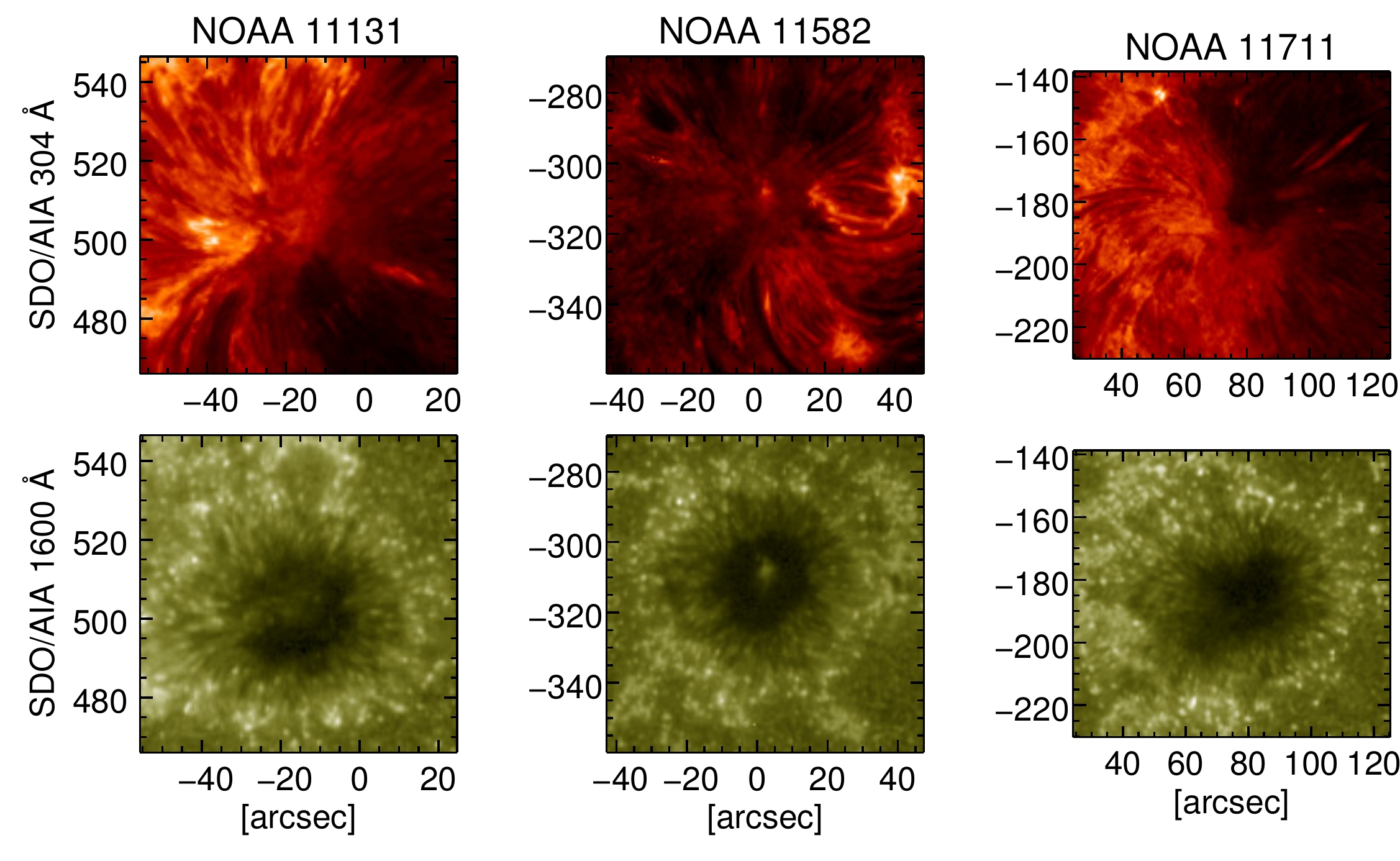}}
  	\caption{Images of the active regions investigated observed at two different wavelengths}
  	\label{fig:fig2}
  \end{figure}

  For the illustration of the technique developed, we selected three active regions: NOAA 11131 (8~December 2010 from 05\,00 until 15\,00 UT), NOAA 11582 (2~October 2012, 05\,00 until 15\,00 UT), and NOAA 11711 (6~April 2013 from 05\,00 until 15\,00 UT). Every  selected active region contains a large sunspot. The observation times were selected near the sunspot passage through the central meridian. For each active region, we used ten hours of SDO/AIA observations  at the wavelengths of 1600\,\AA~ and  304\,\AA\ (see Figure \ref{fig:fig2}). We used the highest available cadence of 12 seconds for the EUV channel and 24 seconds for the UV channel. Cropped and derotated  images were downloaded from the SDO data processing centre web-page  \url{jsoc.stanford.edu}.

\begin{figure} 
  	\centerline{\includegraphics[width=1.\textwidth,clip=]{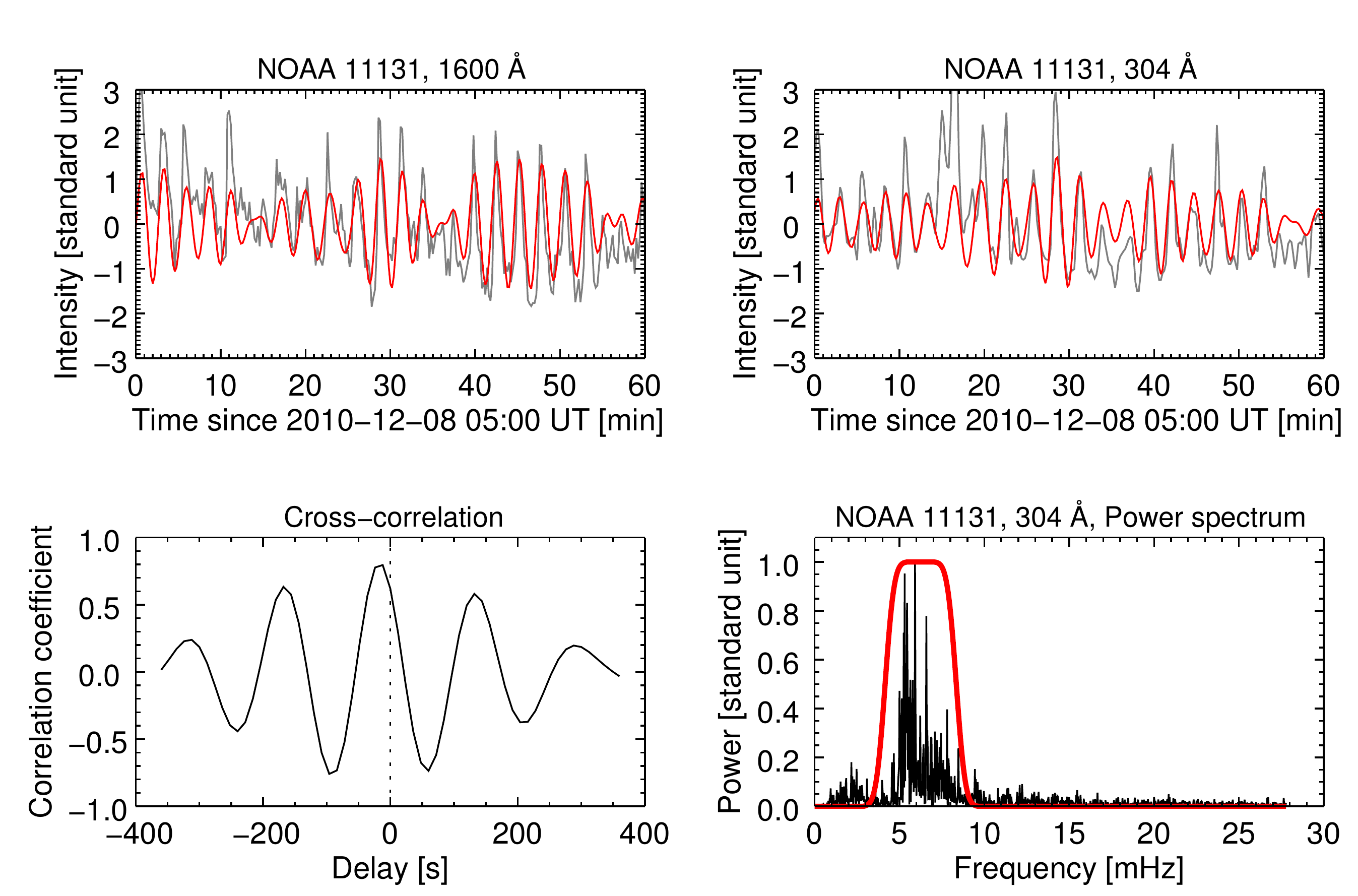}}
  	\caption{Intensity  signals at  the central point of the sunspot umbrae of the active region  NOAA 11131 in 1600\,\AA\ (upper-left panel) and 304\,\AA\ (upper-right panel) SDO/AIA channels, the corresponding cross-correlation function (bottom-left panel) and the normalised power spectrum over-plotted with the band-pass filter used to extract the three-minutes oscillations (bottom-right panel).The original signals are shown by the grey lines, while the thick red lines indicate the bandpass filtering results.  To make individual oscillation cycles clearly visible, we plotted only the first hour of the observational datasets, while the cross-correlation function and the power spectrum are computed for the whole of the observational interval (ten hours)}
  	\label{fig:fig7}
  \end{figure}

  Normally, SDO/AIA observations are evenly spaced in time. However, gaps in the image sequence are possible.
  Furthermore, the cadence and image registration times are different for the selected SDO/AIA channels.
  Therefore, we interpolated observational data obtained in both 1600 and 304\,\AA\ channels to the same instant of time with the constant cadence of 12 seconds. The oscillatory component with the periods ranging from two to four minutes  was extracted from the original signal with the use of Fourier bandpass filtering with a smoothed rectangular window. The intensity variations in the 1600 and 304\,\AA\ channels  extracted from the central pixel of the sunspot NOAA 11131 and the cross-correlation function are shown in Figure \ref{fig:fig7}.
  
  For each active region we downloaded a vector magnetogram observed with the Heliospheric Magnetic Imager (SDO/HMI) at the time corresponding to the  first EUV image in a data set. The magnetic field was extrapolated to the height of 500\,km using the non-linear force free field extrapolation code developed by \cite{2009SoPh..257..287R}.
  
\begin{figure} 
  	\centerline{\includegraphics[width=1\textwidth,clip=]{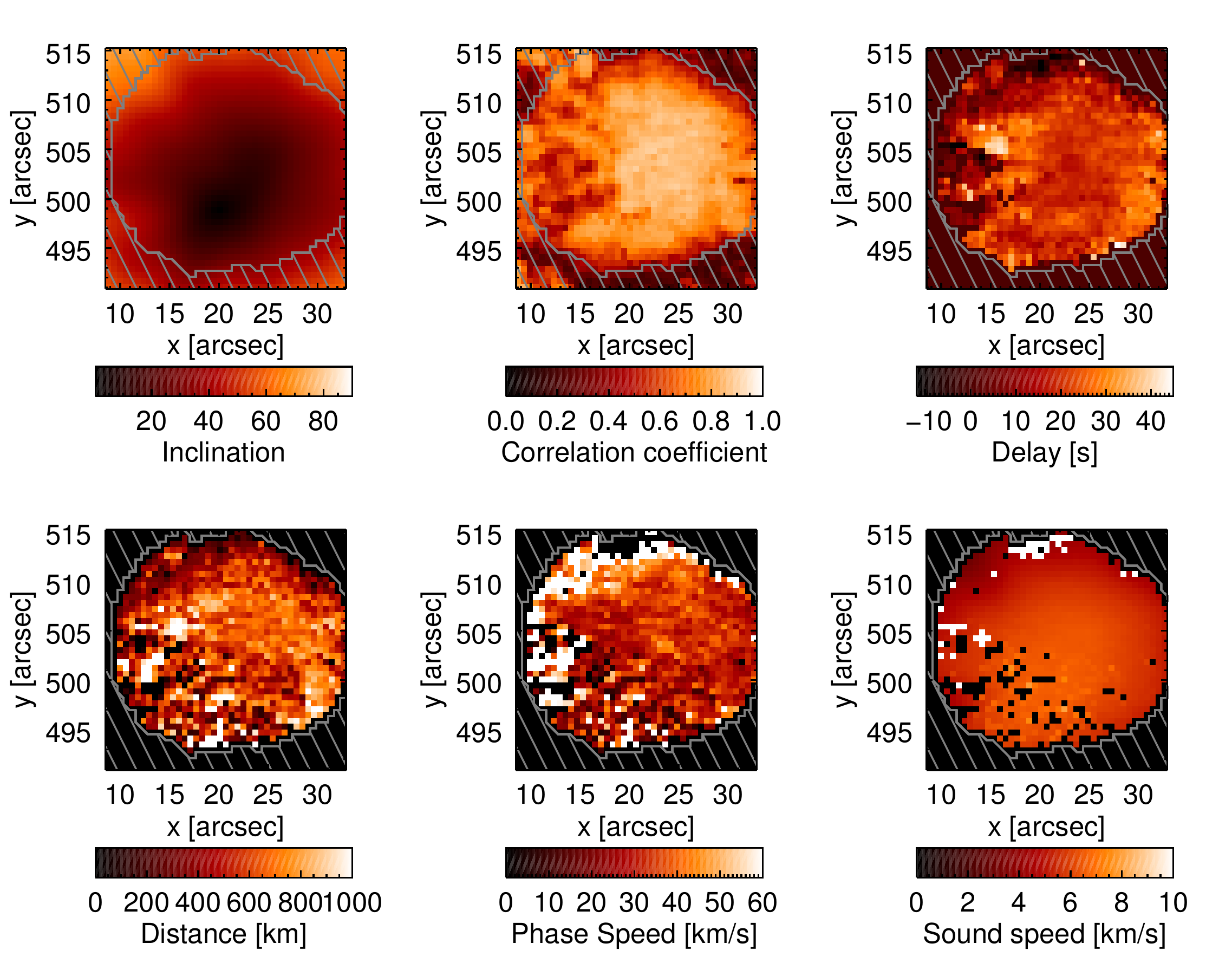}}
  	\caption{Inclination of the magnetic field with respect to the vertical direction, cross correlation coefficient, delay, distance, and phase and sound speeds, calculated for each pixel for active region NOAA 11131}   
  	\label{fig:fig8}
  \end{figure}

\begin{figure} 
  	\centerline{\includegraphics[width = 1\textwidth,clip=]{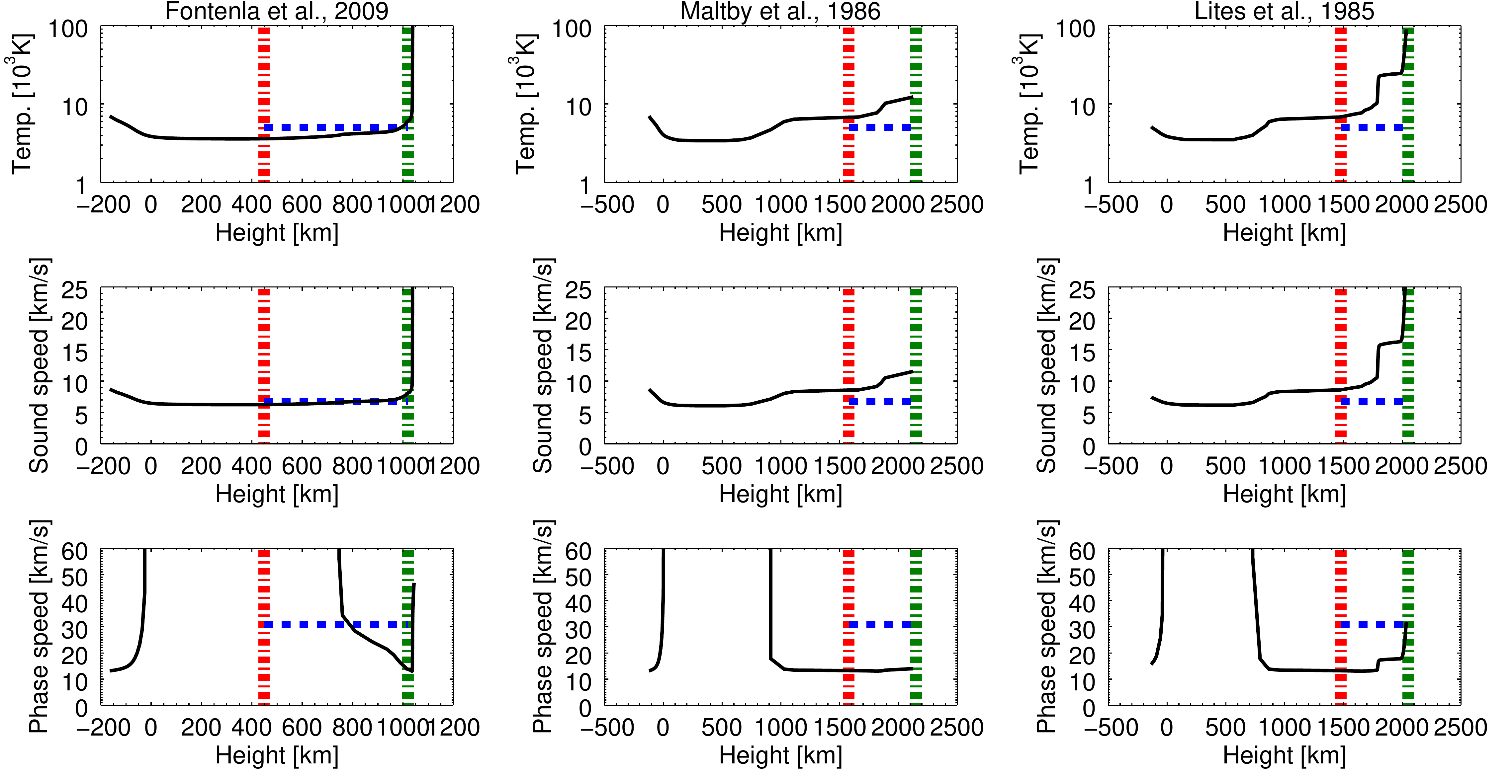}}
  	\caption{Comparison of the seismological estimation of the distance between the layers corresponding to the 1600\,\AA\ and 304\,\AA\ SDO/AIA channels with three models of the sunspot umbrae atmosphere \citep{fontenla_semiempirical_2009,maltby_new_1986, 1985ApJ...294..682L}. The 304~\AA\  layer (vertical-green line) was put either at the height where the temperature reaches 50,000\,K in the model or at the top of the model. While the height of the 1600\,\AA\ layer (vertical-red line) is determined by the distance between the layers inferred for the active region NOAA 11131 (see Table \ref{tbl:1}). The horizontal-blue line shows our seismological estimation of the phase speed, sound speed and the temperature}
  	\label{fig:models}
  \end{figure}

  We analysed the observational data using the method described in Section~\ref{sec: method}. For each active region we obtained the spatial distribution of the distance between the layers and the phase speed. An example  is given in Figure  \ref{fig:fig8}, where  the resulting maps of the measured correlation coefficient, phase delay, estimated distance between the layers and the phase seed for the active region NOAA 11131 are shown. Note that the spatial distribution  of the phase delay is not completely uniform and resembles the distribution of the inferred distance between the layers, despite the both quantities are estimated independently.  This similarity is expected because the wave needs more time to cover a longer distance, assuming the average propagation speed to be the same.
  
The average distance between the temperature minimum (1600\,\AA) and the transition region (304\,\AA), phase delay and phase speed in active regions NOAA 11131, 11582, and 11711 are presented in Table \ref{tbl:1}. 

 \begin{table}
 \caption{Measured parameters calculated for the slow magnetoacoustic wave travelling from the temperature minimum (1600\,\AA) to the transition region (304\,\AA)}\label{tbl:1}

 \begin{tabular}{cccccc}     
 \hline 
 NOAA &Dist. [km] &Delay [s] &Ph. sp. [\,km\,s$^{-1}$] &S. sp. [\,km\,s$^{-1}$] &Temp. [K] \\ 
 \hline
  11131 &$572\pm176$ &$20\pm4$ &$31\pm7$ &$6\pm1.0$ &$3300\pm1000$ \\
  11582 &$749\pm243$ &$25\pm10$ &$30\pm10$ &$5\pm0.7$ &$2300\pm700$ \\
  11711 &$670\pm320$ &$24\pm6$ &$29\pm9$ &$6\pm0.3$ &$3300\pm300$ \\ 
 \hline
 \label{tbl:results}
 \end{tabular}
 \end{table}
 
We found that the distance between the temperature minimum (1600\,\AA) and the transition region (304\,\AA) lies in the range of 500 -- 800\,km for the sunspot umbrae.
The estimated phase speed was found to be about 30\,km\,s$^{-1}$.

\section{Results and Conclusion}
\label{sec:conclusion}

We present a new seismological method allowing for the estimation of the vertical distance, average phase, sound speeds, and temperature between two levels of the sunspot atmosphere observed at different wavelengths. The average temperature is estimated from the average sound speed assuming weakly ionised atmosphere ($\mu = 1.26\,m_h$) and $\gamma =5/3$.

The method uses the three-minute oscillations as a probe  travelling along the magnetic-field lines from the lower levels of the solar atmosphere upwards to the corona. It utilises the assumption that the waves are freely propagating, i.e. the effects connected with the possible partial reflection and transverse structuring were neglected.
The proposed algorithm  uses  image sequences corresponding to two different levels of the sunspot atmosphere and the photospheric magnetic-field measurements as an input.
It does not rely on any assumption about the speed of the wave propagation or spectral-line formation heights.

We applied the designed technique to analysing UV and EUV observations of three active regions [NOAA 11131, 11582, and 1711] and estimated the average sound speed, temperature, and the distance between the temperature minimum (1600\,\AA) and transition region (304\,\AA) above the sunspot umbrae. The estimated distance between temperature minimum and transition region lies in the range of 500 -- 800\,km. The inferred temperatures lie in the range of 2300 -- 4300\,K (see Table \ref{tbl:results}) and roughly correspond to the temperature minimum in the sunspot umbrae.
 
 We compared our measurements with three models of  sunspot umbrae atmosphere \citep{fontenla_semiempirical_2009,maltby_new_1986, 1985ApJ...294..682L}. Figure \ref{fig:models} shows the dependence of the temperature, sound speed and the phase speed of the three-minute. MAG waves upon the height, over-plotted with the inferred positions of the layers visible in  1600\,\AA\ and 304\,\AA\ channels, and average values of the temperature, sound speed, and phase speed between them.
 Figure \ref{fig:models} clearly demonstrates that our results agree with the most recent and advanced semiempirical model of the sunspot umbrae atmosphere  \citep[model S]{fontenla_semiempirical_2009}, which implies a sharp increase in the temperature at the height of about 1000\,km above the photosphere from around 3500\,K  up to the coronal values.
Two other models \citep{maltby_new_1986, 1985ApJ...294..682L} are not consistent with our measurements because they predict the existence of a 1000~km  wide chromosphere between the temperature minimum and the transition region, which is confirmed by neither distance nor temperature  estimations.

 \begin{acks}
  This work was supported by the Russian Foundation for Basic Research, projects:  16-32-00315 mol\_a and 15-02-03835 a. We also thank NASA/SDO, AIA and HMI  teams.
 \end{acks}
 
 \begin{acks}[Disclosure of Potential Conflicts of Interest]
 The authors declare that they have no conflicts of interest.
\end{acks}

 \bibliographystyle{spr-mp-sola}
 
 \bibliography{paper_eng}  
 
\end{article} 
\end{document}